\newif\ifsnjnl
\IfFileExists{sn-jnl.cls}{\snjnltrue}{\snjnlfalse}
\ifsnjnl
  \documentclass[pdflatex,sn-mathphys-num]{sn-jnl}
\else
  \documentclass[12pt]{article}
  \usepackage[margin=25mm]{geometry}
  \usepackage{authblk}
  \newcommand{\email}[1]{\texttt{#1}}
  \newcommand{\bmhead}[1]{\paragraph*{#1}}
  \newcommand{\backmatter}{}
  \renewcommand{\abstract}[1]{\begin{center}\begin{minipage}{0.9\textwidth}\small\textbf{Abstract.} #1\end{minipage}\end{center}}
  \newcommand{\keywords}[1]{\begin{center}\begin{minipage}{0.9\textwidth}\small\textbf{Keywords:} #1\end{minipage}\end{center}}
\fi

\usepackage{amsmath,amssymb}
\usepackage{graphicx}
\usepackage{booktabs}
\usepackage{multirow}
\providecommand{\botrule}{\bottomrule}
\usepackage{xcolor}
\usepackage{pgfplots}
\pgfplotsset{compat=1.17}
\usepackage{hyperref}

\makeatletter
\newcommand{\pu}{\,p.u.\@ifnextchar.{\@gobble}{}}
\makeatother

\begin{document}
\raggedbottom
\title[ Efficient Edge AI for Power Systems ]{Towards Efficient Edge AI for Power Flow in
Renewable-Rich Grids with Volt/VAR Control}

%VS Ver 1: Toward Efficient Edge AI for Power Systems with Volt/VAR Control
% VS Ver 2: Distilling Volt/VAR-Aware GridFM for Edge Applications

%% Author order to be confirmed by the co-authors
\ifsnjnl
\author*[1]{Shahariya R.\ R.\ Puthanpurayil}\email{srrp@energy.aau.dk}
\author[2]{Vlad A. Sanduleac}\email{vlad.sanduleac@upb.ro}
\author*[1]{Remus Teodorescu}\email{ret@energy.aau.dk}
\else
\author[1]{Shahariya R.\ R.\ Puthanpurayil (corresponding, srrp@energy.aau.dk)}
\author[2]{Vlad Sanduleac}
\author[1]{Remus Teodorescu (corresponding, ret@energy.aau.dk)}
\fi

\affil[1]{AAU Energy, Aalborg University, Pontoppidanstræde 111, 9220 Aalborg, Denmark}
\affil[2]{National University of Science and Technology POLITEHNICA Bucharest, Bucharest, Romania}

\ifsnjnl\else\date{}\maketitle\fi

%% Abstract: < 250 words (journal requirement)
\abstract{AI models of the power grid must capture how wind and solar plants support the voltage,
adjusting their reactive power to the local voltage along a grid-code Q(V) characteristic. This
research examines whether grid foundation models, fast neural network models for power-flow analysis, can represent this behaviour. We introduce the \emph{Q(V) bus}, a new bus type for the
open-source GridFM GraphKit, whose physics decoder evaluates the reactive injection from the plant
characteristic at the predicted voltage in every correction step. Training data comprise 7\,882
hourly operating points of a modified IEEE 118-bus system, matched to Energinet DK1/DK2 measurements
and solved with pandapower. Compared with a physics-informed baseline that represents the parks as
negative load, the Q(V) bus lowers the voltage-magnitude RMSE 2.3-fold, reduces the power-balance
mismatch at the parks from 25 to 0.66\,MVA, and raises the F1 score for voltage-band violations from
0.40 to 0.76. An ablation over five configurations shows that the Q(V) bus type provides the accuracy
and the physics loss , the power balance. For edge monitoring, we distil the trained  $2.0\times10^{7}$-parameter
model into students of $6\times10^{5}$ and $1.3\times10^{6}$ parameters; a flat-start initialisation
prevents their collapse into an incorrect low-voltage state in which the characteristic saturates.
With output distillation, the 12-step student reaches 1.4 times the teacher's error and, for edge inference on an NVIDIA
Jetson Orin Nano (15\,W), solves an hourly scenario in 12.2\,ms, within the 20\,ms synchrophasor interval.}


\ifsnjnl\maketitle\fi

%% =====================================================================
\section{Introduction}\label{sec:intro}

Converter-interfaced wind and solar plants now supply a large share of the electricity in systems
such as the Danish grid, and they also take part in voltage control. Grid codes require these plants
to adjust their reactive power to the local voltage according to a volt/VAR, or Q(V), characteristic
\cite{ncrfg,energinet_rfg,ieee1547}. A power-flow (PF) model of such a grid must therefore treat the
reactive power of a renewable plant as a function of its own voltage, not as a fixed value.

Graph neural networks are a natural representation of a power grid \cite{liao2022}, and grid
foundation models (GridFMs) extend this idea to graph transformers pretrained on a masked PF
objective with a Kirchhoff physics loss \cite{hamann2024,puech2025}. While implementing two gaps remain for
renewable-rich grids. First, the bus types in GridFM follow the classical PQ, PV and slack
convention, so a Q(V)-controlled plant can only be represented as a PQ bus, whose reactive power is
either given, although it is unknown in operation, or masked. Second, the models are sized for
data-centre inference, whereas monitoring of low-inertia grids increasingly asks for inference next
to the measurement, within one synchrophasor reporting frame of 20--100\,ms
\cite{milano2018,ieeec37118,gooi2023}. In earlier work we compressed a GridFM by knowledge
distillation (KD) into a 32.5$\times$ smaller student for an NVIDIA Jetson device and found that
output matching carried most of the gain \cite{suanduleac2026physics}; that study used only the classical bus
types.

This paper addresses both gaps. We add a Q(V) bus type to GridFM, train it on hourly Danish system
data, and compress it into small students for edge inference. The contributions are:
\begin{itemize}
  \item An element-aware Q(V) bus type for GridFM, in which the reactive power of a park follows its
        Q(V) characteristic at the predicted voltage, and an ablation over five teacher configurations showing that the Q(V) bus type provides the
accuracy and the physics loss the power balance, which requires the correct reactive reference
that the characteristic supplies at the parks.

  \item A dataset of 7\,882 hourly scenarios built from measured Danish consumption and production
        (Energinet, DK1/DK2, October 2025 to September 2026) on a modified IEEE 118-bus grid with
        32 Q(V)-controlled parks, solved with pandapower \cite{thurner2018} and written in the GridFM
        DataKit format \cite{puech2025}.
  \item  Students of the Q(V) model (0.6\,M and 1.3\,M parameters): a flat-start initialisation for
        reliable training, an evaluation of output and trajectory distillation, and a latency and
        throughput evaluation on an NVIDIA Jetson Orin Nano against a Newton--Raphson solver with
        Q(V) control.
\end{itemize}

Section~\ref{sec:related} reviews related work, Section~\ref{sec:methods} describes the grid, data
and models, Section~\ref{sec:results} presents the teacher, student and edge-inference results,
Section~\ref{sec:discussion} discusses their implications and limitations, and
Section~\ref{sec:conclusions} concludes.

%% ============================================================introduction what is wrong there =========
\section{Related work}\label{sec:related}

\paragraph{Learning-based power flow and grid foundation models.}
The first graph neural solver for AC PF was trained by minimising the Kirchhoff violation directly \cite{donon2020}; physics-guided variants followed for PF \cite{hu2021}, state estimation \cite{ngo2024} and optimal power flow \cite{nellikkath2022}. Message-passing surrogates such as PowerFlowNet report speed-ups of 4--48$\times$ over Newton--Raphson \cite{lin2024}, while generalisation across topologies remains open \cite{arowolo2025}. GridFMs address this by pretraining on many topologies and operating points \cite{hamann2024,puech2025};their power-flow model keeps the graph-neural-solver idea of refining the state over several correction steps with physics residuals. These models use the classical bus types; to our knowledge, none represents voltage-dependent reactive control of renewable plants as a bus type of its own.

\paragraph{Volt/VAR control of renewable plants.}
IEEE Std 1547-2018 defines volt/VAR functions for distributed resources, in which the reactive power follows a piecewise-linear function of the local voltage, saturating at the reactive limits outside a voltage band \cite{ieee1547}; transmission-connected plants follow similar characteristics set by the national grid codes \cite{ncrfg,energinet_rfg}. In conventional PF tools such characteristics are handled by an outer loop around the Newton--Raphson iteration \cite{thurner2018}. A learned PF model has to include the characteristic in its prediction to learn from, which changes both the masking (what is known and unknown at a bus) and the physics loss.

\paragraph{Knowledge distillation and compact models for the edge.}
KD trains a small student to reproduce a larger teacher through its outputs, intermediate states or attention maps \cite{hinton2015,romero2015,gou2021}, and has been extended to graph networks \cite{tian2025}. Student accuracy and fidelity to the teacher are known to diverge \cite{stanton2021}. Edge intelligence for power systems is an active field \cite{gooi2023}, and for graph networks the number of message-passing layers dominates the latency on embedded devices \cite{barillaro2025}. Our earlier study distilled a GridFM into a half-depth student on a renewable IEEE 118-bus benchmark and measured it on a Jetson device \cite{suanduleac2026physics}; it used only the classical bus types.

%% =====================================================================
\section{Grid, data and models}\label{sec:methods}

\subsection{DK118 grid and Energinet scenarios}\label{sec:data}

We use a modified IEEE 118-bus grid onto which hourly Energinet measurements of consumption and of wind, solar and conventional production in the Danish bidding zones DK1 and DK2 are mapped~\cite{energinet}. The 118 buses are split into two connected areas of 59 buses, a synthetic representation of DK1 and DK2, and all active-power quantities are scaled by one common factor that maps the largest measured Danish gross consumption to the nominal IEEE 118-bus demand of 4\,242\,MW. The 54 generator buses of the IEEE case are assigned to technologies in the proportions of the Danish transmission system, which has a high renewable share: 32 wind and solar power park modules (hereafter \emph{parks}) are modelled as Q(V) buses, and 20 synchronous units and 2 synchronous condensers as PV buses (Table~\ref{tab:sites}), with larger units placed on more strongly connected buses. Wind and solar parks are listed separately in Table~\ref{tab:sites} because they follow different production profiles; however ,while in data generation and learning models they are treated alike as parks in Q(V) control. In every hour, the wind and solar production of each zone equals its scaled Energinet value and is shared among the parks of that zone. Most units are connected at 138\,kV;
four parks are connected to the 345\,kV network.

Each hourly scenario is solved with pandapower~\cite{thurner2018}, which serves as ground truth. The lines, transformers and shunts of the IEEE case are converted with pandapower's MATPOWER converter,and the scaled loads and injections are added for every hour: synchronous units as
voltage-controlled generators with reactive limits, the slack unit as an external grid, HVDC exchanges as fixed injections, and wind and solar parks as static generators whose reactive power follows their Q(V) characteristic. Because pandapower's built-in Q(V) controller, a damped
fixed-point loop, diverges for these steep characteristics, the park injections are instead updated by a Newton step around pandapower's power flow; synchronous units switch between PV and PQ at their reactive limits in both directions, and the zonal AC losses are kept within the measured
consumption. Voltage set-points and the centres of the Q(V) characteristics are initially set by a loss-minimising optimal power flow over training days only. The period from 7 October 2025 to 1 September 2026 (330 days) gives 7\,898 complete hours, of which 16 do not converge and are
excluded, leaving 7\,882 scenarios in the GridFM DataKit format~\cite{puech2025,gridfm_datakit_github}. The scenarios are split by whole days, assigned at random, into training, validation and test sets (230/50/50 days, about 70/15/15\,\%), with 5\,489, 1\,197 and 1\,196 hours.

\begin{table}[h]
\caption{Renewable-dominated generation fleet placed on the IEEE 118-bus grid (IEEE bus numbers)}\label{tab:sites}%
\begin{tabular*}{\textwidth}{@{\extracolsep\fill}lp{11.5pc}p{11.5pc}@{}}
\toprule
Unit type & DK1 & DK2 \\
\midrule
Wind, Q(V)      & 1, 4, 6, 8, 12, 18, 19, 34, 36, 46, 49, 61, 65 & 25, 73, 80, 87, 89, 100, 111 \\
Solar, Q(V)     & 10, 15, 26, 40, 42, 54, 59 & 24, 27, 31, 32, 103 \\
Synchronous, PV & 55, 56, 62, 66 & 69\footnotemark[1], 70, 72, 74, 76, 77, 85, 90, 91, 92, 99, 104, 105, 107, 110, 112 \\
Condensers, PV  & -- & 113, 116 \\
\botrule
\end{tabular*}
\footnotetext[1]{Slack (reference) bus.}
\end{table}

\subsection{ The Q(V) bus: an element-aware bus type}\label{sec:qv}
In the masked PF task of GridFM, the unknown quantities at a bus depend on its type: PQ buses predict $(V_m, V_a)$, PV buses predict $(V_a, Q_g)$ and the slack bus predicts $(P_g, Q_g, V_m)$
\cite{hamann2024,gridfm_graphkit_github}. A PQ bus is normally a load bus, and a unit with a fixed output can be represented
on it as an opposite injection. Starting with this,we design a  baseline models that represents each
renewable park in this way: its known active power enters as negative load. Its reactive power, however, is unknown in operation under Q(V) control and is therefore masked;the existing GridFM decoder then sets it to zero (Section~\ref{sec:teachers}).

The proposed model, GridFM-Q(V), The proposed model, GridFM-Q(V), keeps the architecture and training procedure of GridFM as base with modifications making each bus aware of the element connected to it. It adds a fourth bus type, the Q(V) bus:At a Q(V) bus, the active power is known and $(V_m, V_a)$ are predicted; the reactive power is not
an input but follows from the characteristic, $Q_g = q(V_m)$.In existing architecture , the park's PQ flag is cleared, an indicator marks it as Q(V), and its active power is given as generation. In addition, the physics decoder evaluates the park's reactive power from its Q(V) characteristic at the predicted voltage in every correction step, while the network itself receives only the indicator;the curve parameters come from the case data.

\begin{figure}[t]
  \centering
  \includegraphics{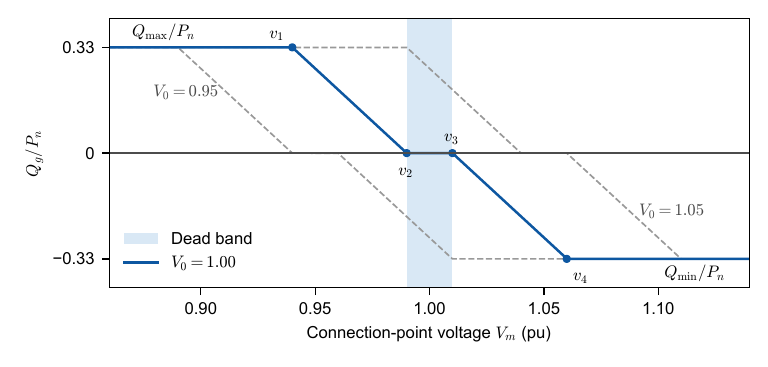}
  \caption{Q(V) characteristic normalised by rated power $P_n$. Solid: $V_0=1.00$\,pu with
  breakpoints $v_1$--$v_4$ and dead band (shaded). Dashed: $V_0=0.95$ and $1.05$\,pu, the range
  of curve centres in the data.}
  \label{fig:qv}
\end{figure}

Each park follows the same Q(V) characteristic, shifted to its own curve centre $V_0$
(Fig.~\ref{fig:qv}). With breakpoints $(v_1, v_2, v_3, v_4) = V_0 + (-0.06,\,-0.01,\,+0.01,\,+0.06)$\,pu,
\begin{equation}
  \begin{aligned}
    Q_g = q(V_m) &= Q_{\max}\,\mathrm{clip}\!\left(\frac{v_2 - V_m}{v_2 - v_1}\right)
                  + Q_{\min}\,\mathrm{clip}\!\left(\frac{V_m - v_3}{v_4 - v_3}\right),\\
    \mathrm{clip}(x) &= \min\bigl(1,\max(0,x)\bigr),
  \end{aligned}
  \label{eq:qv}
\end{equation}
so the park injects reactive power below a dead band of $\pm0.01$\,pu around $V_0$, absorbs above
it, and saturates at its limits. Here $Q_{\max}=-Q_{\min}=0.33\,P_n$, with $P_n$ the rated power of
the park, and $V_0$ is fixed per park between 0.95 and 1.05\,pu by the voltage schedule
(Section~\ref{sec:data}). These settings lie within the NC RfG voltage control requirements for
power park modules~\cite{ncrfg,energinet_rfg}; the breakpoints themselves are a study
assumption.

GridFM-Q(V) thus differs from existing GridFM in two respects: the Q(V) bus type, with the characteristic
in the physics decoder, and the physics loss evaluated with this characteristic. An ablation study
separates their effects; it defines the teacher configurations of Section~\ref{sec:teachers}, and
its results are analysed in Section~\ref{sec:results}.

\subsection{Teacher models and ablation design}\label{sec:teachers}
The teacher is a heterogeneous GridFM graph transformer with 12 message-passing steps, hidden size 48 and 8 attention heads (20.1\,M parameters), trained with GridFM GraphKit. The loss combines the masked mean-squared error of weight 0.9 with the nodal power-balance residual of weight 0.1; the physics term weights step $k$ of an $L$-step model by $0.5^{L-k}$. Five configurations, each trained with three
seeds, form the ablation study (Table~\ref{tab:configs}).

In GridFM-Base1 and GridFM-Base2, each park is treated as an ordinary PQ bus, with its active power as negative load and its reactive power masked. Since a physics residual is only meaningful for known injections, GridFM-Base1 retains only the active-power residual at the park buses, as GridFM does for the unknown reactive power of PV buses. GridFM-Base2 is the existing GridFM without modification, which evaluates the residual for both active and reactive power with the masked park reactive power set to zero; it is kept as the reference that a direct application of GridFM would give. The proposed GridFM-Q(V) uses the Q(V) bus and evaluates the park's reactive power with Eq.~\eqref{eq:qv} in the residual. GridFM-MSE and GridFM-Q(V)-MSE are the corresponding models trained without any of the physics loss for baseline and proposed models respectively.

\begin{table}[h]
\caption{Teacher configurations (three seeds each). Park $Q_g$ in the correction steps: reactive power of
the parks used in the power-balance residual that the model computes in every step; physics loss: the
part of this residual included in the training loss}\label{tab:configs}%
\begin{tabular}{@{}llll@{}}
\toprule
Configuration & Bus-type aware & Park $Q_g$ in the correction steps & Physics loss at park buses \\
\midrule
GridFM-Base1     & no  & from the nodal balance & $P$ only \\
GridFM-Base2     & no  & $Q_g = 0$              & $P$ and $Q$ \\
GridFM-Q(V)      & yes & $Q_g = q(V_m)$         & $P$ and $Q$ \\
GridFM-MSE       & no  & $Q_g = 0$              & none \\
GridFM-Q(V)-MSE  & yes & $Q_g = q(V_m)$         & none \\
\botrule
\end{tabular}
\end{table}

All teachers are trained for 200 epochs with AdamW (learning rate $5\times10^{-4}$, batch size 64),
a linear warm-up over five epochs and cosine decay to $10^{-6}$, keeping the weights with the lowest
validation loss. In a preliminary comparison, a reduce-on-plateau schedule (200 and 400 epochs)
lowered the learning rate prematurely and was therefore not used.

\subsection{Student models and distillation}\label{sec:students}

Students keep the teacher's input interface and loss, with hidden size 24 and 4 attention heads (96 numbers per bus per step, against 384 for the teacher). We use a 6-step student (0.62,M parameters, 32.5$\times$ smaller) and a 12-step student (1.31,M parameters, 15.3$\times$ smaller). Following our earlier finding that the output term carries nearly all of the distillation gain \cite{suanduleac2026physics}, students are trained without KD, with output distillation, in which the
teacher's predictions act as a second label next to the pandapower solution, or with output plus
trajectory distillation. We set $\lambda_{\mathrm{out}}=0.75$ and add this weighted output-distillation term to the supervised and physics losses. The teacher's parameters remain frozen during student training. Base1 students of the same sizes are trained with and without output
KD from the GridFM-Base1 teacher as a reference. %Stock PF students of the same sizes are trained in the same way from the GridFM\_PF teacher as a reference.
%\todo{Verify the last sentence above}

Specifically, the training objective is
% \begin{equation}
% \label{eq:total-distillation-loss}
% \mathcal{L}_{\mathrm{total}}
% =
% \mathcal{L}_{\mathrm{GraphKit}}
% +
% \lambda_{\mathrm{out}}\mathcal{L}_{\mathrm{out}},
% \end{equation}
\begin{equation}
\label{eq:total-distillation-loss}
\mathcal{L}_{\mathrm{total}}
=
\mathcal{L}_{\mathrm{GraphKit}}
+
\lambda_{\mathrm{out}}\mathcal{L}_{\mathrm{out}}
+
\lambda_{\mathrm{traj}}\mathcal{L}_{\mathrm{traj}}.
\end{equation}
where $\mathcal{L}_{\mathrm{GraphKit}}
=0.9\,\mathcal{L}_{\mathrm{sup}}
+0.1\,\mathcal{L}_{\mathrm{phys}}$ combines the supervised voltage
loss against the pandapower solution with the physics loss over
the correction steps. The output-distillation term is
\begin{equation}
\label{eq:output-distillation-loss}
\begin{aligned}
\mathcal{L}_{\mathrm{out}}
=\frac{1}{2}\Bigg[
&\frac{1}{|\mathcal{M}_b|}
\sum_{(i,d)\in\mathcal{M}_b}
\left(
\hat{y}^{\,b}_{S,i,d}-\hat{y}^{\,b}_{T,i,d}
\right)^2
\\
+{}&
\frac{1}{|\mathcal{M}_g|}
\sum_{j\in\mathcal{M}_g}
\left(
\hat{y}^{\,g}_{S,j}-\hat{y}^{\,g}_{T,j}
\right)^2
\Bigg].
\end{aligned}
\end{equation}
Here, $S$ and $T$ denote the student and teacher, respectively.
The set $\mathcal{M}_b$ contains the masked voltage-magnitude and
angle entries, and $\mathcal{M}_g$ contains the masked generator
active-power entries in the minibatch. The predictions $\hat{y}$
use the training normalisation. The factor $1/2$ averages the
bus and generator matching losses.

All final students use flat-start initialisation: the weights of the
voltage-magnitude output row are initialised to zero and its bias to
1\,pu. These parameters remain trainable.

We compare no KD, output KD, and output plus trajectory KD, with
$(\lambda_{\mathrm{out}},\lambda_{\mathrm{traj}})$ equal to
$(0,0)$, $(0.75,0)$ and $(0.75,0.75)$, respectively.
The trajectory loss is the mean masked MSE of voltage magnitude and
angle across all student correction steps, including the final step.
Student step $k$ matches teacher step
$m(k)=\operatorname{round}(kL_T/L_S)$, where $L_T=12$ and $L_S$
is the student depth. Thus six-step students match teacher steps
2, 4, 6, 8, 10 and 12; twelve-step students match corresponding steps.
The teacher remains frozen and the exact Q(V) curve is retained.

\subsection{Evaluation metrics}\label{sec:metrics}
All models are evaluated on the 1\,196 test hours and reported as mean $\pm$ standard deviation
over three seeds. Accuracy is the root-mean-square error (RMSE) of the quantities that are unknown
at each bus type: $V_m$ and $V_a$ at PQ and Q(V) buses, $V_a$ at PV buses and $V_m$ at the
reference bus. The unknown
generator injections follow from the predicted voltages through the power balance or the Q(V)
characteristic and are not scored separately. Whether the predicted voltages satisfy the power-flow equations is measured by the power-balance
mismatch (PBE), the mean nodal apparent-power mismatch of the predicted voltages over the known
injections: active power at all buses except the reference bus, reactive power at PQ buses and
$q(V_m)$ at Q(V) buses. The PBE uses the true device behaviour for every model, so no model is judged
by its own decoder. The operational value is measured by the precision, recall and F1 score of
detecting buses outside the 0.98--1.04\,pu operating band. As a no-skill reference, each bus's $V_m$
and $V_a$ are predicted by their mean over the training hours.

%% =====================================================================
\section{Results}\label{sec:results}

\subsection{Teacher models: voltage accuracy and power balance}\label{sec:res-teachers}

\begin{figure}[h]
  \centering
  \includegraphics{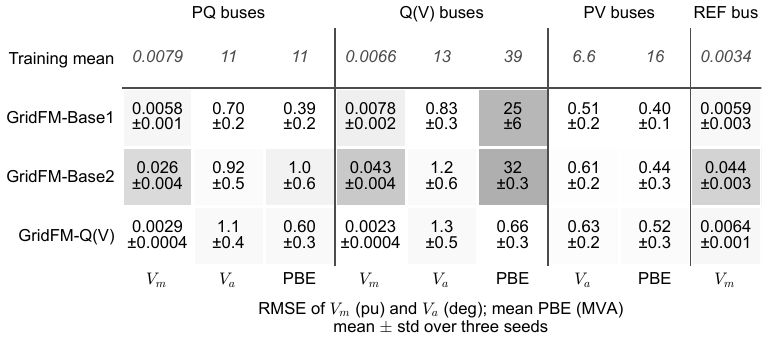}
  \caption{Test errors of the teachers by bus type: RMSE of $V_m$ and $V_a$ and mean PBE
  (1\,196 test hours, mean $\pm$ standard deviation over three seeds). Each quantity is evaluated
  only where it is unknown; the PBE is evaluated for the known injections. Shading: error relative
  to the lowest error in each column, including the training-mean reference (darker = larger).}
  \label{fig:heatmap}
\end{figure}

\begin{figure}[h]
  \centering
  \includegraphics{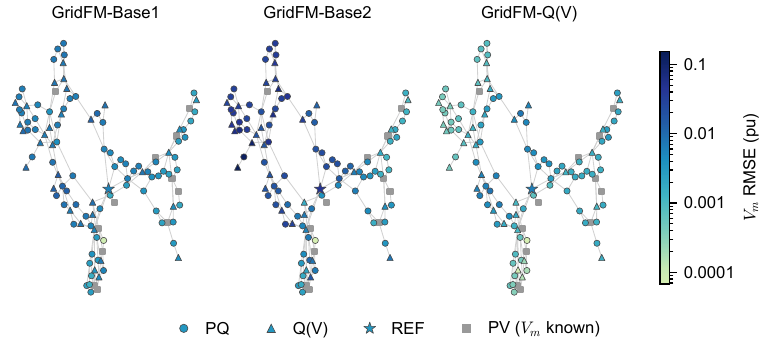}
  \caption{Voltage-magnitude RMSE per bus on the grid layout over the test hours, averaged over three
  seeds. PV buses (grey) have a known voltage magnitude and are not evaluated.}
  \label{fig:busmap}
\end{figure}

The teacher comparison tests the central hypothesis of this work: that representing renewable parks
by their Q(V) control behaviour, rather than as negative load, yields voltages that are both more
accurate and consistent with the power-flow equations. GridFM-Q(V) confirms this (Fig.~\ref{fig:heatmap}): it reaches a
voltage-magnitude RMSE of 0.0029\,pu at PQ buses and 0.0023\,pu at the parks, closes the power
balance at the parks to 0.66\,MVA and predicts their reactive power with an RMSE of 1.9\,Mvar. Its
errors are low across the grid (Fig.~\ref{fig:busmap}) and concentrated in few hours: its RMSE is
2.6 times its mean absolute error, against 1.4--1.7 for the other teachers, and the largest
deviations, up to 0.07\,pu, occur in the hours with the heaviest load and the lowest voltages, which
are rare in the training data.Over all buses where it is unknown, its voltage-magnitude RMSE is 0.0028\,pu, 2.3 times lower than that
of GridFM-Base1 (0.0065\,pu).

GridFM-Base1, which keeps the parks as negative load but restricts the physics residual to the known
injections, is the stronger of the two physics-informed baselines. It gives more accurate angles than
GridFM-Q(V) (RMSE 0.5--0.8$^\circ$, against 0.6--1.3$^\circ$), but its voltage-magnitude RMSE is 2.0
times higher at PQ buses (0.0058\,pu) and 3.4 times higher at the parks (0.0078\,pu), where it remains
above the training mean (0.0066\,pu). At PQ and PV buses, both models satisfy the power balance closely
(0.4--0.6\,MVA), but at the parks that of GridFM-Base1 is violated by 25\,MVA and their reactive power
is off by 30\,Mvar: without an equation for the park's reactive power, the predicted voltages are
consistent with some reactive injection, but not with the Q(V) characteristic. Eq.~\eqref{eq:qv}
supplies this missing equation. For both models, the voltage of the reference bus, released when the
slack unit reaches its reactive limit, lies above the training mean (0.0059 and 0.0064\,pu against
0.0034\,pu); for GridFM-Q(V) it is the largest voltage error

The existing GridFM residual (GridFM-Base2) additionally checks the reactive balance at the parks
against $Q_g=0$. This incorrect reference raises the voltage-magnitude RMSE 4.5 to 7.5 times
relative to GridFM-Base1, above the training mean at every bus type, while the angles change much
less: it distorts the reactive power-voltage relation, most strongly around the 345\,kV parks in
DK1 (Fig.~\ref{fig:busmap}).

The voltage errors translate directly into operational value (Table~\ref{tab:f1}). GridFM-Base2
raises mostly false band-violation alarms (precision 0.07), as its biased voltages move compliant
buses out of the band, whereas GridFM-Base1 detects only 28\,\% of the violations. GridFM-Q(V)
combines a precision of 0.94 with a recall of 0.64 (F1 0.76): its alarms can be trusted, and it
finds more than twice as many violations as GridFM-Base1.

\begin{table}[h]
\caption{Detection of buses outside the 0.98--1.04\,pu band on the test set (5\,038 violations;
mean $\pm$ standard deviation over three seeds)}\label{tab:f1}%
\begin{tabular}{@{}lccc@{}}
\toprule
Model & Precision & Recall & F1 \\
\midrule
Training mean & 0.85            & 0.20            & 0.33 \\
GridFM-Base2  & $0.07 \pm 0.01$ & $0.58 \pm 0.05$ & $0.13 \pm 0.01$ \\
GridFM-Base1  & $0.69 \pm 0.15$ & $0.28 \pm 0.01$ & $0.40 \pm 0.03$ \\
GridFM-Q(V)   & $0.94 \pm 0.01$ & $0.64 \pm 0.04$ & $0.76 \pm 0.03$ \\
\botrule
\end{tabular}
\end{table}

\subsection{Student Models and Distillation}\label{sec:res-students}
The students test whether the accuracy of GridFM-Q(V) can be retained in a model small enough for
edge deployment, and whether knowledge distillation helps. All students have hidden size 24 and
4 heads, with 12 steps (1.3\,M parameters) or 6 steps (0.6\,M), against 20.1\,M for the teacher.

% Without further measures, the Q(V) students collapse: their first voltage estimate in the 345\,kV
% region falls below the sloped part of the Q(V) characteristic, where the park reactive power is
% clamped at its maximum and the characteristic provides no correction signal or gradient, so the
% student settles in a self-consistent wrong state. This affected all 6-step and three of six
% 12-step students. Initialising the first voltage estimate at 1.0\,pu (flat start), as numerical
% power-flow solvers do, removes the collapse: none of the 12 flat-start students collapsed.

\begin{table}[h]
% \caption{Q(V) students with flat start: voltage-magnitude error over all buses (pu, test set,
% mean $\pm$ standard deviation over three seeds); teacher 0.0011\,pu}\label{tab:students}%
\caption{Voltage-magnitude RMSE (pu) over unknown outputs for Q(V) and
Base1 students on the test set. Values are mean $\pm$ standard deviation
over three seeds. All students use flat-start initialisation. The
corresponding teacher RMSEs are 0.002795\,pu for Q(V) and 0.006487\,pu
for Base1. Dashes indicate variants that were not evaluated.}
\begin{tabular}{@{}lccc@{}}
\toprule
Student & No KD & Output KD & Output + trajectory KD \\
\midrule
% 12 steps (1.3\,M) & $0.0021 \pm 0.0003$ & $0.0018 \pm 0.0006$ & \todo{X} \\
% 6 steps (0.6\,M)  & $0.0029 \pm 0.0001$ & \todo{X}           & $0.0034 \pm 0.0003$ \\
% Q(V), 12 steps (1.3M) & $0.0021 \pm 0.0003$ & $0.0018 \pm 0.0006$ & $0.0017 \pm 0.0004$ \\
% Q(V), 6 steps (0.6M)  & $0.0029 \pm 0.0001$ & $0.0031 \pm 0.0001$ & $0.0034 \pm 0.0003$ \\
% \midrule
% Base1, 12 steps (1.3M) & $0.0067 \pm 0.0003$ & $0.0067 \pm 0.0010$ & --- \\
% Base1, 6 steps (0.6M)  & $0.0089 \pm 0.0006$ & $0.0080 \pm 0.0008$ & --- \\
Q(V), 12 steps (1.3M) & $0.0045 \pm 0.0004$ & $0.0040 \pm 0.0009$ & $0.0037 \pm 0.0006$ \\
Q(V), 6 steps (0.6M)  & $0.0054 \pm 0.0001$ & $0.0057 \pm 0.0002$ & $0.0061 \pm 0.0004$ \\
\midrule
Base1, 12 steps (1.3M) & $0.0090 \pm 0.0002$ & $0.0090 \pm 0.0011$ & --- \\
Base1, 6 steps (0.6M)  & $0.0114 \pm 0.0006$ & $0.0104 \pm 0.0009$ & --- \\
\botrule
\end{tabular}
\end{table}

% \todo{Accuracy paragraph: students reach about twice the teacher's error at 12 steps; output KD
% lowers the 12-step error by about 13\,\% on average and improves the angles in all seeds; effect at
% 6 steps; trajectory KD.}

At twelve steps, output KD reduces mean voltage-magnitude RMSE by
10.6\,\% and mean PBE from 0.835 to 0.658\,MVA compared with no KD.
Angle error improves in all three matched seeds, while voltage-magnitude
error improves in two. Adding trajectory KD further lowers mean
voltage-magnitude RMSE to 0.003718\,pu, but increases mean PBE to
0.754\,MVA and angle RMSE from 1.065 to 1.207 degrees.

At six steps, neither KD variant improves mean voltage-magnitude
error or PBE, although both reduce mean angle error. The benefit of
distillation therefore depends on student depth. The best twelve-step
voltage result remains about 1.3 times the teacher's error, at
15.3 times fewer parameters.

% \todo{Base1 students, one paragraph: whether the compact GridFM-Base1 remains behind the compact
% GridFM-Q(V), as for the teachers.}
The effect of output KD differs between the two model families.
For twelve-step Base1 students, mean voltage-magnitude RMSE  remains
approximately 0.0090\,pu, while its standard deviation across seeds
increases from 0.0002 to 0.0011\,pu. Mean PBE nevertheless decreases
from 12.074 to 11.421\,MVA. The Q(V) students retain lower absolute
voltage-magnitude and balance errors, although their advantage is
also present without KD. Comparing the three-seed mean RMSEs, the
twelve-step output-KD student has 42.6\,\% higher voltage-magnitude
error than its teacher for Q(V), compared with 39.3\,\% for Base1.
Thus the more accurate Q(V) student still has a larger relative gap
to its stronger teacher. At six steps, output KD reduces Base1's
voltage-magnitude RMSE by 8.5\,\%, whereas it increases the Q(V)
student's error by 5.5\,\%. Both families improve their mean angle
accuracy at this depth.

Table~\ref{tab:student-angle-balance} compares the angle and
power-balance errors across student depths and distillation settings.

\begin{table}[tbp]
\centering
\small
\caption{Student voltage-angle RMSE and mean PBE on the test set,
reported as mean $\pm$ standard deviation over three seeds.}
\label{tab:student-angle-balance}
\begin{tabular}{llcc}
\toprule
Student & Training & Angle RMSE ($^\circ$) & Mean PBE (MVA) \\
\midrule
Q(V), 12 steps
 & No KD
 & $1.298 \pm 0.054$ & $0.835 \pm 0.163$ \\
 & Output KD
 & $1.065 \pm 0.099$ & $0.658 \pm 0.202$ \\
 & Output + trajectory KD
 & $1.207 \pm 0.468$ & $0.754 \pm 0.276$ \\
\midrule
Q(V), 6 steps
 & No KD
 & $1.633 \pm 0.052$ & $1.715 \pm 0.125$ \\
 & Output KD
 & $1.465 \pm 0.048$ & $1.763 \pm 0.060$ \\
 & Output + trajectory KD
 & $1.503 \pm 0.051$ & $1.790 \pm 0.172$ \\
\midrule
Base1, 12 steps
 & No KD
 & $0.957 \pm 0.165$ & $12.074 \pm 0.515$ \\
 & Output KD
 & $0.921 \pm 0.375$ & $11.421 \pm 1.430$ \\
\midrule
Base1, 6 steps
 & No KD
 & $1.486 \pm 0.261$ & $12.242 \pm 0.847$ \\
 & Output KD
 & $1.277 \pm 0.257$ & $11.946 \pm 0.498$ \\
\bottomrule
\end{tabular}
\end{table}

% Figure~\ref{fig:student-angle-balance} compares the angle and
% power-balance errors across student depths and distillation settings.

% \todo{Determine if these two charts should remain here to help with the discussion/explanation}
% \begin{figure}[tbp]
% \centering
% \includegraphics[width=\linewidth]{charts/student_Va.png}
% \par\medskip
% \includegraphics[width=\linewidth]{charts/student_balance_S.png}
% \caption{Voltage-angle MAE (top) and mean PBE (bottom) for
% twelve-step (left) and six-step (right) students.
% Results show mean $\pm$ standard deviation over three seeds;
% dashed lines indicate teacher means.}
% \label{fig:student-angle-balance}
% \end{figure}

%% =====================================================================
\subsection{Edge inference}\label{sec:edge}

The models were deployed on an NVIDIA Jetson Orin Nano Developer Kit, a 15\,W embedded module that
combines a six-core Arm CPU with an integrated Ampere GPU (1\,024 CUDA cores, 32 Tensor Cores) and
8\,GB of memory shared by both. The module ran in its 15\,W power mode with fixed clocks (CPU 1.5\,GHz,
GPU 625\,MHz; JetPack 5.1.3, PyTorch 2.1.0, CUDA 11.4), and the desktop environment was stopped during
all measurements, since screen redraws share the integrated GPU. Each hourly scenario is one
power-flow problem, so latency is measured at batch size one, as in online monitoring, over all
1\,196 test scenarios after 50 warm-up runs; throughput is measured with batches of up to 128
scenarios. One trained seed of each model is deployed in float32 and reproduces its stored test
predictions to within $2\times10^{-6}$\,pu.

At batch size one, a forward pass consists of about 200 small GPU kernels per correction step. In
standard execution, the CPU launches each kernel separately, and this kernel-launch overhead would
dominate the run time. We therefore employ a CUDA graph: the complete forward pass, including the Q(V)
physics step, is recorded once and replayed for every scenario with a single launch, and the inputs of
the scenario reach the GPU in one host-to-device copy. This aligns with the normal power-flow operation
considered here: the branch projections are computed once, while the bus types, setpoints, loads,
injections and Q(V) parameters are copied in per scenario. The timed region contains the copy and one
replay.

The classical reference is a Newton--Raphson power flow with each park's $q(V_m)$ in the mismatch and
its slope in the Jacobian, built on pandapower's Jacobian routines. It receives the same inputs as the
models, runs from a flat start ($V_m=1$\,pu, $V_a=0$) and from the solution of the preceding hourly
scenario (warm start), and agrees with the dataset to within $3\times10^{-8}$\,pu. The solve is timed
on one CPU core; the six-core run serves as a throughput reference.

\begin{table}[h]
\caption{Latency per scenario on the Jetson Orin Nano (float32, batch size one, 1\,196 test scenarios)
and best throughput in scenarios per second. NR: Newton--Raphson. Speed-up: mean latency of the Q(V)
NR solver from a flat start divided by that of the method. Throughput of the models at batch size 128;
classical solvers on one CPU core, except the last row}\label{tab:latency}%
\begin{tabular*}{\textwidth}{@{\extracolsep\fill}lrrrrr@{}}
\toprule
 & \multicolumn{3}{@{}c@{}}{Latency (ms)} & & \\
\cmidrule{2-4}
Method & Mean & Median & p99 & Speed-up & Throughput \\
\midrule
Teacher, 12 steps (20.1\,M)  & 24.59 & 24.59 & 24.64 & 1.4 & 73 \\
Student, 12 steps (1.3\,M)   & 12.18 & 12.18 & 12.21 & 2.9 & 355 \\
Student, 6 steps (0.6\,M)    &  6.16 &  6.16 &  6.19 & 5.6 & 710 \\
\midrule
Q(V) NR, flat start          & 34.77 & 31.22 & 46.82 & 1.0 & 29 \\
Q(V) NR, warm start          & 24.34 & 23.31 & 31.18 & 1.4 & 41 \\
Q(V) NR, 6 CPU cores         & --    & --    & --    & --  & 152 \\
\botrule
\end{tabular*}
\end{table}

\paragraph{Latency and throughput.}
On the Jetson, the 12-step student solves a scenario in 12.2\,ms and the 6-step student in 6.2\,ms,
2.6 and 5.1 times faster than the Q(V) Newton--Raphson solver at the median and also faster than its
warm start (Table~\ref{tab:latency}). The teacher (24.6\,ms) is faster than the flat start but slower
than the warm start, and exceeds the 20\,ms reporting interval of a 50\,Hz synchrophasor stream. The
latency of the models is deterministic (99th percentile within 0.5\,\% of the median), whereas the
varying iteration count of Newton--Raphson (4.5 on average, up to 7) places its 99th percentile
50\,\% above its median. In batches, the 12-step student processes 355 scenarios per second, 2.3 times
as many as Newton--Raphson on all six cores, while the teacher (73 scenarios/s) stays below it. Latency
follows the number of correction steps rather than the number of parameters: halving the steps halves
the student's latency, and the teacher, four times wider than the 12-step student with the same number
of GPU operations, is twice as slow. Since the deployed models run in float32, their accuracy is that
of Section~\ref{sec:res-students}; the 12-step student, with 1.4 times the teacher's voltage error, is
therefore the edge model of choice, whereas the 6-step student detects too few band violations for
monitoring. We acknowledge that in a trial run on a workstation CPU (AMD Ryzen Threadripper PRO
7955WX, 16 cores), Newton--Raphson solved a scenario nearly twice as fast as the 12-step student on
the Jetson. However, this work focuses on energy-efficient inference at the edge, and the workstation
solver required about 2.6 times the energy per solution of the 12-step student on the Jetson.

\section{Discussion}\label{sec:discussion}

\paragraph{Bus type and physics loss.}
The study examines the suitability of a  physics loss and how it improves the power balance of a learned power-flow model. With existing bus representation of a park as a PQ bus with $Q_g=0$,
a reactive injection the park does not produce, and this wrong reference degrades the voltages
(Sect.~\ref{sec:res-teachers}) as in GridFM-Base2. Leaving $Q_g$ free, as in GridFM-Base1, removes the error but
treats the park as an unconstrained reactive source: its reactive power follows from the nodal
balance as at a PV bus, only the active-power equation remains, and the voltage magnitude at the
park is determined by the data alone. Under Q(V) control, the reactive power of a park is neither
fixed, as at a PQ bus, nor unconstrained, as at a PV bus; in steady state, the plant controller
sets it from the local voltage within its reactive limits. The Q(V) characteristic therefore
defines the reactive power of the park through its control law, which the numerical power flow
used to generate the data also enforces, and provides the reactive-power equation at the park bus.

Table~\ref{tab:ablation} separates the contributions of the bus type and the physics loss. The Q(V)
bus type gives the lowest voltage-magnitude errors of all teachers, with and without the physics
loss (0.0023\,pu without and 0.0028\,pu with the physics loss, against 0.0050 and 0.0065\,pu for the
PQ representation)
, and the
highest F1 scores. Without the physics loss, however, the predicted voltages leave a power-balance
mismatch of 15\,MVA on average and 19\,MVA at the parks. The physics loss reduces this mismatch to
below 1\,MVA at a slightly higher voltage-magnitude RMSE, so that GridFM-Q(V) is the only model that combines a low voltage error with a satisfied power
balance among the studied ones.

\begin{table}[h]
\caption{Ablation of park representation and physics loss (test set, mean over three seeds). RMSE
of $V_m$ (pu) and $V_a$ ($^\circ$) over all buses where they are unknown, and of the angle
difference across the branches, $\Delta V_a$ ($^\circ$); PBE: mean power-balance mismatch (MVA)
over all buses and at the parks}\label{tab:ablation}%
\begin{tabular*}{\textwidth}{@{\extracolsep\fill}lcccccccc@{}}
\toprule
 & & & \multicolumn{3}{@{}c@{}}{RMSE} & \multicolumn{2}{@{}c@{}}{PBE} & \\
\cmidrule{4-6}\cmidrule{7-8}
Model & Park bus & Physics loss & $V_m$ & $V_a$ & $\Delta V_a$ & All & Parks & F1 \\
\midrule
GridFM-MSE      & PQ   & no  & 0.0050 & 0.76 & 0.52 & 35   & 42   & 0.43 \\
GridFM-Base1    & PQ   & yes & 0.0065 & 0.72 & 0.13 & 7.1  & 25   & 0.40 \\
GridFM-Q(V)-MSE & Q(V) & no  & 0.0023 & 0.35 & 0.22 & 15   & 19   & 0.73 \\
GridFM-Q(V)     & Q(V) & yes & 0.0028 & 1.12 & 0.17 & 0.61 & 0.66 & 0.76 \\
\botrule
\end{tabular*}
\end{table}

\paragraph{Voltage angle.}
As we reviewed the result ,the voltage angle error of GridFM-Q(V) have different indication. Its voltage angles are less accurate than
those of GridFM-Q(V)-MSE and GridFM-Base1 (RMSE 1.12$^\circ$ against 0.35 and 0.72$^\circ$, and
0.7-1.5$^\circ$ across seeds), but the angle differences across the branches, which set the
active-power flows, at least as accurate as without the physics loss (0.17 against 0.22$^\circ$). The
error therefore lies in the angles relative to the reference bus and does not affect the flows or
the power balance. Its cause requires further investigation.

\paragraph{Student models -Distillation.}
% The richer bus model changes what a student model has to learn: its first voltage estimate must
% already lie on the sloped part of the Q(V) characteristic, where the characteristic provides a
% correction signal. Initialising this estimate at 1.0\,pu, as numerical solvers do, was sufficient to
% train all compact students reliably. Output distillation gave a modest gain at 12 steps
% \todo{and at 6 steps}, and \todo{trajectory distillation}. Since a physically inconsistent
% prediction shows up in its power-balance error without ground truth, this error can serve as an
% online check, with a fall-back to a numerical solver when it exceeds a threshold. \todo{Jetson
% latency (p50/p99, batch size 1) of the teacher and the students.}

The benefit of distillation depends on student depth and on the
quantity being predicted. For the twelve-step Q(V) student, output
KD improves mean voltage-magnitude, angle and power-balance errors.
For the six-step Q(V) student, it reduces angle RMSE from 1.633 to
1.465 degrees, but increases voltage-magnitude RMSE from 0.005427
to 0.005726\,pu. Mean PBE also increases slightly, from 1.715 to
1.763\,MVA. These results show that the tested distillation loss
does not improve every output of the smaller PF model; they do
not establish that it has reached a capacity limit.

Trajectory KD gives a further voltage-magnitude improvement only
for the twelve-step Q(V) student. Relative to output KD, it reduces
mean voltage-magnitude RMSE by 6.7\,\%, but increases angle RMSE
from 1.065 to 1.207 degrees and mean PBE from 0.658 to 0.754\,MVA.
It also increases voltage-magnitude error at six steps. The extra
trajectory term therefore changes the balance between the evaluated
quantities, rather than providing a consistent improvement across
them. Its benefit in these experiments is the lower mean
voltage-magnitude error of the twelve-step student.

PBE can be evaluated from the predicted state, known injections,
and the supplied network and device models without a reference
PF solution. It therefore provides an online check of the power balance during inference.
 Its use to trigger a numerical-solver
fallback was not evaluated in this study.

% \paragraph{Depth and width of the students.}
% Depth and width play different roles. Each message-passing step acts like one correction of an
% iterative solver, so depth sets how far the estimate can be refined: with flat start, the 12-step
% student (1.3\,M parameters) reaches a voltage-magnitude error of 0.0021\,pu against 0.0029\,pu for
% the 6-step student (0.6\,M), at twice the inference cost. Width, in contrast, decides whether the
% first estimate is good enough to start from. In a 60-epoch screening without flat start, students
% with hidden size 48 did not collapse (0 of 3 seeds, with 4 or 8 heads), whereas hidden size 24
% collapsed in 2 of 3 seeds even with 8 heads, so the per-step width rather than the number of heads
% matters. Flat start removes this dependence on width, which allows the narrow and therefore fast
% students used here. \todo{Effect of distillation at 6 versus 12 steps.}

\paragraph{Student depth.}

Each GridFM correction step combines attention-weighted message
passing over the graph connections with voltage prediction and
power-balance residual feedback. Attention weights messages from
connected neighbours. Additional steps therefore allow information
from more distant nodes to influence the predictions and provide
further stages of refinement. For Q(V) students with output KD,
increasing depth from six to twelve steps reduces mean
voltage-magnitude RMSE by 30.4\,\%, angle RMSE by 27.3\,\% and mean
PBE by 62.7\,\%; operating-band F1 increases from 0.224 to 0.532.
However, parameter count also increases from 0.62M to 1.31M, so
this comparison does not isolate information propagation from
additional model capacity. The final students have the same hidden
size and number of heads, and therefore do not establish the effect
of width. Additional steps increase sequential computation; their
latency cost is measured in Section~\ref{sec:edge}.

\paragraph{Performance across the network.}

The Q(V) voltage-magnitude advantage extends beyond the renewable
parks, but is larger at the RE buses. For twelve-step output-KD
students, mean voltage-magnitude RMSE at RE buses decreases from
0.010696\,pu for Base1 to 0.003017\,pu for Q(V). At non-RE buses,
it decreases from 0.008205 to 0.004345\,pu. The ordering between
the two bus groups also reverses: Base1 has larger voltage errors
at RE buses than elsewhere, whereas Q(V) has smaller errors there.
The balance improvement is concentrated at the parks, where mean
PBE is 40.368\,MVA for Base1 and 0.711\,MVA for Q(V). At the other
eligible buses, Base1 has the lower mean PBE, 0.524 versus
0.638\,MVA. Both models have larger mean PBE at RE buses than
elsewhere, but this comparison does not establish the cause of
that difference.

\paragraph{Operating-band detection.}

At twelve steps, the Q(V) students retain higher operating-band
F1 than their Base1 counterparts, but remain below the Q(V) teacher.
For the output-KD student, precision decreases from the teacher's
0.939 to 0.891, while recall decreases more substantially, from
0.636 to 0.385. With trajectory KD, precision is 0.918 and recall
is 0.399. Thus most generated alarms still identify genuinely
out-of-band buses, but many actual out-of-band bus-hours are missed.
At six steps, the comparison changes: the output-KD Q(V) student
has higher precision than Base1 (0.682 versus 0.592), but much lower
recall (0.134 versus 0.300), resulting in lower F1 (0.224 versus
0.398). Lower voltage-magnitude RMSE therefore does not consistently
translate into better operating-band detection. These scores concern
the 0.98--1.04\,pu operating band, not the hard voltage limits.

\paragraph{Inference optimisation.}
Floating-point formats trade resolution for speed. We first applied float16 to the whole network,
with weights and intermediate values in float16 and only the physics step in float32. Float16
resolves voltages near 1\,pu only to about $10^{-3}$\,pu, the size of the teacher's error: the output
heads round the voltages, and the hidden state accumulates rounding errors over the correction steps.
This raised the teacher's voltage error by 20\,\%, far beyond the 1\,\% we accepted relative to
float32, and bfloat16, with an even coarser resolution, failed in the same way. We then applied mixed
precision (FP16-mixed), as in automatic mixed precision: only the query, key, value and skip
projections, which hold nearly all of the arithmetic, run in float16, with their inputs cast to
float16 and their outputs back to float32, while attention, normalisation, the hidden state, the
output heads and the physics step stay in float32. With FP16-mixed, the errors stay within 0.2\,\% of
float32 for all models (Table~\ref{tab:precision}), and the teacher becomes 1.36 times faster (24.6 to
18.2\,ms), so that the most accurate model fits within the 20\,ms synchrophasor interval.

\begin{table}[h]
\caption{FP16-mixed (query, key, value and skip projections in float16, everything else in float32)
against float32 on the Jetson Orin Nano: latency per scenario (batch size one), gain in throughput,
and relative change of the voltage-magnitude and angle errors over all test scenarios at the buses
where they are unknown ($+$: larger error)}\label{tab:precision}%
\begin{tabular*}{\textwidth}{@{\extracolsep\fill}lrrrrrr@{}}
\toprule
 & \multicolumn{2}{@{}c@{}}{Latency (ms)} & & & \multicolumn{2}{@{}c@{}}{Error change (\%)} \\
\cmidrule{2-3}\cmidrule{6-7}
Model & float32 & FP16-mixed & Speed-up & Throughput & $V_m$ & $V_a$ \\
\midrule
Teacher, 12 steps  & 24.59 & 18.15 & 1.36 & 1.37 & $0.0$  & $+0.1$ \\
Student, 12 steps  & 12.18 & 12.00 & 1.02 & 1.08 & $-0.1$ & $-0.1$ \\
Student, 6 steps   &  6.16 &  6.09 & 1.01 & 1.07 & $-0.1$ & $+0.2$ \\
\botrule
\end{tabular*}
\end{table}

For the students, FP16-mixed shortens the latency by only 1--2\,\%, because their run time is set by
the roughly 200 small GPU operations per correction step, whose number does not depend on the width,
rather than by arithmetic. 
\paragraph{Limitations.}
The study has several limitations. All parks follow one generalised Q(V) characteristic, whereas in
practice settings differ between parks and other control modes are used; HVDC links are represented
as fixed injections without the voltage support of their converters. The fleet is a synthetic
placement of Danish production on the IEEE 118-bus grid, and the data cover normal operation only,
without faults, outages or other abnormal events. The voltage of the reference bus, released when
the slack unit reaches its reactive limit, remains the largest voltage error of GridFM-Q(V).
Robustness to changed curve settings, park locations and topologies requires further
out-of-distribution tests. The physics and data terms of the loss use fixed weights
(0.1/0.9) that were not tuned for the Q(V) bus type.Finally,the edge deployment assumes the fixed topology of normal operation; line outages or switching actions
would require a CUDA graph per topology, so the reported latencies are a lower bound for operation under
contingencies.

%% =====================================================================
\section{Conclusions and future work}\label{sec:conclusions}

We introduced a Q(V) bus type for GridFM, in which the reactive power of a renewable park follows its
Q(V) characteristic at the predicted voltage, and trained it on measured Danish hourly data on a
modified IEEE 118-bus grid. Compared with a physics-informed baseline that represents the parks as
negative load, it lowers the voltage-magnitude RMSE 2.3-fold, reduces the power-balance mismatch at
the parks from 25 to 0.66\,MVA and raises the F1 score for voltage-band violations from 0.40 to 0.76.
The ablation shows that the bus type provides the accuracy and the physics loss the power balance.
Distilled students of the Q(V) model collapse without further measures, as train reliably with a flat
start. With output distillation, the 12-step student reaches 1.4 times the teacher's error and solves
a scenario on an NVIDIA Jetson Orin Nano in 12.2\,ms, about three times faster than a Newton--Raphson
solver with Q(V) control and within the 20\,ms synchrophasor interval; with mixed precision, the
teacher also meets this interval.

Several questions remain open. The bus model should be extended to individual curve settings and other
control modes, and its behaviour under topology changes, abnormal operating points and the voltage
support of HVDC converters remains to be investigated. The cause of the angle error relative to the
reference bus and its dependence on the weighting of the physics loss require further investigation.
Methods and edge devices that further reduce the latency of the students should be explored, and the
power-balance mismatch remains to be evaluated as an online trigger for a fall-back to a numerical
solver.

%% =====================================================================
\backmatter

\section*{Declarations}

\bmhead{Funding}
This work was supported by the Villum Foundation through the Villum Investigator project Smart Battery
(project no.~222860) at Aalborg University.

\bmhead{Competing interests}
The authors declare no competing interests.

\bmhead{Ethics approval and consent to participate}
Not applicable.

\bmhead{Clinical trial number}
Not applicable.

\bmhead{Consent for publication}
Not applicable.

\bmhead{Data availability}
The hourly consumption and production data are publicly available from Energinet's Energi Data Service.

\cite{energinet}. The derived grid dataset and the trained models are available from the corresponding
author on reasonable request.

\bmhead{Code availability}
The source code and scripts supporting the findings of this study are available from the corresponding
author on reasonable request.

\bmhead{Author contributions}
S.R.R.P.: Conceptualization, Methodology, Investigation, Writing -- original draft, Writing -- review \& editing.
V.S.: Investigation, Writing -- original draft, Writing -- review \& editing.
R.T.: Conceptualization, Writing -- original draft, Writing -- review \& editing.
All authors read and approved the final manuscript.

\bmhead{Use of generative AI}
In line with the Springer Nature policy on large language models (LLMs), we document their use.
An LLM-based coding assistant (Claude Code, Anthropic) was used to write, debug and run the code
for data generation, training and evaluation, to produce figures, to help draft and edit the
manuscript text, and as a discussion aid during the design and analysis of the experiments. All
experimental design decisions and the interpretation of results were made by the authors, who
cross-checked the code, results, figures and text and take full responsibility for the content of
this paper. No LLM is listed as an author.

%% =====================================================================

\bibliography{references}% references.bib in the same folder; no \bibliographystyle needed

\end{document}